\documentclass[twocolumn,aps,prl,amsmath,amssymb]{revtex4-2}
\usepackage{graphicx}
\usepackage{dcolumn}
\usepackage{bm}

\usepackage{multirow}

\usepackage{amssymb}
\usepackage{amsmath}
\usepackage{graphicx}
\usepackage{xcolor}
\usepackage[colorlinks,citecolor=blue,linkcolor=red,urlcolor=blue]{hyperref}
\usepackage{braket}
\usepackage{CJK}
\usepackage{indentfirst}
\usepackage{amsmath}
\usepackage{cases}
\usepackage{enumerate}
\usepackage{booktabs} 
\usepackage{tabularx} 

\def\beq{\begin{equation}}
\def\eeq{\end{equation}}
\def\bald{\begin{aligned}}
\def\eald{\end{aligned}}
\def\bea{\begin{eqnarray}}
\def\eea{\end{eqnarray}}

\def\bra#1{\left\langle#1\right|}
\def\ket#1{\left|#1\right\rangle}
\def\avg#1{\left\langle#1\right\rangle}

\def\Eq#1{Eq.~(\ref{#1})}

\def\bra#1{\left\langle#1\right|}
\def\ket#1{\left|#1\right\rangle}
\def\avg#1{\left\langle#1\right\rangle}

\def\Eq#1{Eq.~(\ref{#1})}

\definecolor{YinshuaiBlue}{RGB}{0,125,255}
\newcommand{\subref}[1]{\textcolor{YinshuaiBlue}{#1}}

\begin{document}
\title{Universal Entanglement Growth along Imaginary Time in Quantum Critical Systems}
\author{Chang-Yu Shen$^{1,2}$}
\author{Shuai Yin$^{3,4}$}
\email{yinsh6@mail.sysu.edu.cn}
\author{Zi-Xiang Li$^{1,2}$}
\email{zixiangli@iphy.ac.cn}

\affiliation{$^1$Beijing National Laboratory for Condensed Matter Physics $\&$ Institute of Physics, Chinese Academy of Sciences, Beijing 100190, China}
\affiliation{$^2$University of Chinese Academy of Sciences, Beijing 100049, China}
\affiliation{$^3$Guangdong Provincial Key Laboratory of Magnetoelectric Physics and Devices, School of Physics, Sun Yat-sen University, Guangzhou 510275, China}
\affiliation{$^4$School of Physics, Sun Yat-sen University, Guangzhou 510275, China}

\date{\today}
\begin{abstract}
Characterizing universal entanglement features in higher-dimensional quantum matter is a central goal of quantum information science and condensed matter physics. While the subleading corner terms in two-dimensional quantum systems encapsulate essential universal information of the underlying conformal field theory, our understanding of these features remains remarkably limited compared to their one-dimensional counterparts. We address this challenge by investigating the entanglement dynamics of fermionic systems along the imaginary-time evolution. We uncover a pioneering non-equilibrium scaling law where the corner entanglement entropy grows linearly with the logarithm of imaginary time, dictated solely by the universality class of the quantum critical point. Through unbiased Quantum Monte Carlo simulations, we verify this scaling in the interacting Gross-Neveu-Yukawa model, demonstrating that universal data can be accurately recovered from the early stages of relaxation. Our findings significantly circumvent the computational bottlenecks inherent in reaching full equilibrium convergence. This work establishes a direct link between the fundamental theory of non-equilibrium critical phenomena and the high-precision determination of universal entanglement properties on both classical and quantum platforms, paving the way for probing the rich entanglement structure of quantum critical systems.
\end{abstract}
\footnotetext[1]{These authors contribute equally to this work.}
\maketitle

\noindent {\bf INTRODUCTION}

Quantum entanglement entropy has become a cornerstone for characterizing the universal and non-local properties of quantum many-body systems. For the ground states of gapped systems, the entanglement entropy typically obeys an ``area law'', scaling with the size of the boundary of a subsystem~\cite{RMPEntanglement}. More intriguingly, deviations from this leading behavior are often universal. For instance, for one-dimensional ($1$D) critical systems described by a conformal field theory (CFT), the area law is violated by a logarithmic term whose prefactor is proportional to the central charge~
\cite{Wilczek1994NPB,1dEE_scaling1,1dEE_scaling2,1dEE_scaling3,Jiang2012NP}. The features of entanglement entropy are more complex and intriguing in higher dimensions. In two-dimensional ($2$D) topologically ordered phases, a subleading constant term—the topological entanglement entropy—directly reveals the nature of the topological order~\cite{kitaev2006prl,levin2006prl,Dong_2008,Flammia2009,Nicolas2016PR,Melko2011NP}. Similarly, $2$D quantum critical points (QCPs) display universal logarithmic corrections, notably arising from sharp corners in the entanglement region, which are tied to the underlying CFT~\cite{1dEE_scaling2,Casini_2009,Casini2007NPB,Kallin2013,Fradkin2006PRL,laflorencie2016pr,Bueno2016,Bueno_2015,Bueno2019,Wessel2014PRB,Helmes2016PRB}. However, in stark contrast to the well-understood 1D case, the rich entanglement structure of higher-dimensional quantum many-body systems remains far less explored and warrants significant investigation. The corner contribution is often swamped by the non-universal area-law term, requiring high-precision calculations that defy analytical treatments and challenge state-of-the-art numerical methods.

In two or higher dimensions,  quantum Monte Carlo (QMC) is particularly well-suited for large-scale simulations due to its intrinsic lack of bias~\cite{AssaadReview,li2019review}.  Nevertheless, a generic and efficient evaluation of entanglement entropy within the QMC framework remains a significant hurdle, stemming from the exponential growth of variance when sampling particular partition functions~\cite{Grover2013PRL,Scalettar2014PRB}. This difficulty is exacerbated when extracting the subleading logarithmic corner term via equilibrium QMC, a task that requires isolating a subtle signal from a noisy background. Despite significant methodological advances~\cite{Hastings2010,Humeniuk2012,McMinis2013,Wang2014PRLEE,InglisMelko2013,Assaad2014PRB,Trebst2016PRE,Assaad2015PRB,Emidio2020PRL,Emidio2024PRL,Jiang2024arXiv,Xu2025NC}, including the development of incremental algorithms~\cite{Emidio2024PRL}, the accurate and efficient determination of the universal coefficient for the corner term, especially in interacting fermionic systems, persists as a major challenge.

Meanwhile, the non-equilibrium dynamics of entanglement represents a new interdisciplinary frontier at the intersection of quantum many-body physics and quantum information. Among various non-equilibrium realizations, the imaginary-time evolution constitutes a cornerstone of computational many-body physics, serving as a robust projection method to isolate ground states from generic initial configurations. Beyond its theoretical and numerical utility, imaginary-time evolution has gained further relevance with the advent of quantum computers and simulators, providing a pathway to preparing ground states on quantum devices—a feat already demonstrated experimentally on quantum hardware~\cite{Chan2020NP,Nishi2021npjQI,Pollmann2021PRXQuantum,Zhang2024PRBimaginarytime}. Recent research into imaginary-time dynamics has yielded breakthroughs across multiple fields, such as illuminating non-equilibrium quantum criticality~\cite{yu2023dirac,Yins2014prb,Yin2014pre,Zeng2025NC,Zeng2025PRB,Yin2022prl,Zhong2021prb,Shu2022prb} and devising QMC techniques that circumvent the sign problem~\cite{Yu2024arXiv}. However, analyses of quantum entanglement have remained largely confined to equilibrium. Consequently, despite its fundamental importance, the dynamics of entanglement during imaginary-time evolution in dimensions greater than one remains a largely unexplored territory.

Herein, we address two fundamental challenges simultaneously by investigating the imaginary-time entanglement dynamics of fermionic quantum many-body systems. First, we uncover a pioneering non-equilibrium universal scaling law for entanglement entropy in $2$D critical systems. Second, leveraging this scaling, we establish a highly efficient protocol to extract universal entanglement properties, significantly circumventing the computational bottlenecks of traditional equilibrium methods.

We focus on the universal dynamics originating from geometric corners in the entangling surface. Drawing on the established equilibrium scaling behavior, we propose that during the early stages of imaginary-time evolution, the corner contribution to the entanglement entropy follows a universal logarithmic scaling with evolution time, governed by a size-independent coefficient dictated by the underlying CFT. We validate this hypothesis through systematic simulations of free Dirac fermions, recovering a coefficient consistent with prior equilibrium results. Crucially, by bypassing the heavy computational burden of full equilibrium convergence,  our approach significantly reduces computational overhead. We demonstrate the versatility of this protocol by extracting universal CFT data at the Gross-Neveu-Yukawa (GNY) QCP~\cite{Rosenstein1993PLB,Herbut2006PRL,Assaad2013prx,li2017nc,Scherer2017PRD,Li2018SA,Lang2019prl,Abolhassan2022PRL,Li2024PRLNon-Hermitian} and within the antiferromagnetic (AFM) phase via sign-free QMC. To conclude, our work establishes a robust framework for probing universal entanglement properties in the short-time regime. These findings not only enrich the fundamental theory of non-equilibrium critical phenomena but also provide a highly efficient numerical protocol for entanglement spectroscopy. Moreover, such non-equilibrium strategies offer a viable route toward enhancing the efficiency of quantum algorithms requiring ground-state preparation across diverse quantum platforms.

\noindent {\bf Universal non-equilibrium scaling of entanglement entropy}

In $2$D quantum critical systems, the entanglement entropy encodes the universal properties of the critical point. Specifically, for critical systems without Fermi surface, the second R\'enyi entropy for a subregion $A$ with linear size $L$ follows the scaling form~\cite{Fradkin2006PRL,Casini2007NPB,TomoyoshiHirata_2007,Bueno2019}:
\begin{equation}
	S^A_2(L) = aL - s_c(\theta) \ln L + \text{const.} + \mathcal{O}(1/L).
	\label{scalingeq}
\end{equation}
Here, $aL$ represents the standard area-law contribution found in generic quantum ground states, where the coefficient $a$ is non-universal and depends on microscopic details. The universal component is the logarithmic correction arising from the sharp corner in region $A$. The angle-dependent coefficient $s_c(\theta)$ is independent of microscopic specifics and is determined solely by the universality class of the quantum critical systems. 

\begin{figure}[!t]
	\centering
	\includegraphics[width=\linewidth]{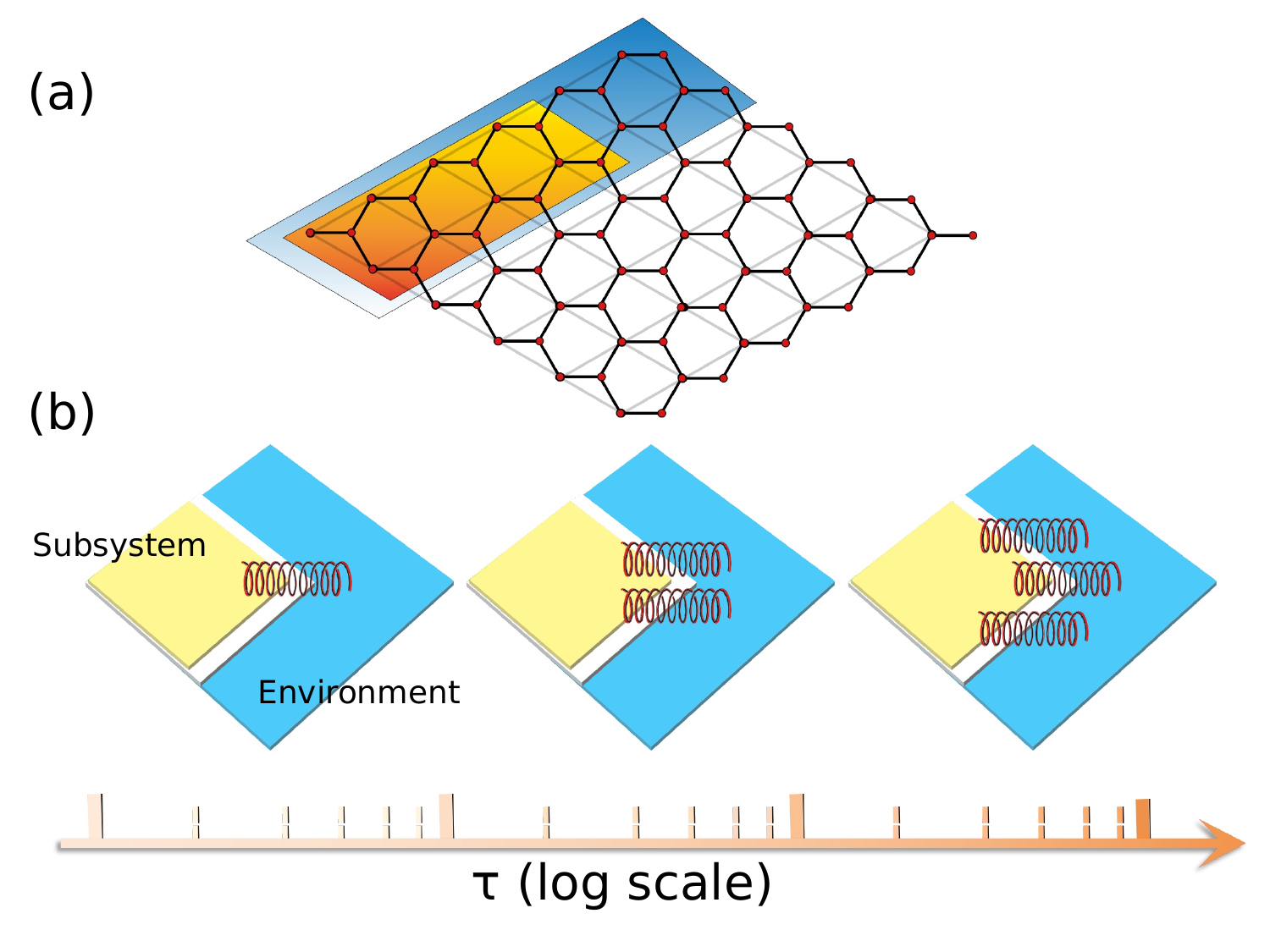}
	\caption{\textbf{(a)} Lattices and EE subsystem cuts considered in this
		Letter. The full system has periodic boundary conditions in
		both horizontal and vertical directions.  The orange region (Region $A$) has dimensions $\frac{L}{3}\times \frac{2L}{3}$ and features two $\frac{\pi}{3}$ and two $\frac{2\pi}{3}$ corners. The blue region (Region $B$) has dimensions $\frac{L}{3}\times L$ and possesses a smooth boundary. Crucially, both regions share the same boundary length.
		\textbf{(b)} Schematic illustration of the corner entanglement growth in imaginary-time evolution. The springs represent the degree of corner contribution to entanglement between the subsystem and the environment. In the non-equilibrium regime, the corner contribution to the entanglement entropy increases linearly with the logarithm of the imaginary time $\tau$.  
	}
	\label{SCEE_png}
\end{figure}

Here, we investigate the scaling behavior of the entanglement entropy during imaginary-time evolution from a product initial stage. In the relaxation process, the wave function $|\psi(\tau)\rangle$ evolves according to the imaginary-time Schr\"{o}dinger equation $-\frac{\partial}{\partial \tau} |\psi(\tau)\rangle=H|\psi(\tau)\rangle$, subject to wavefunction normalization. We focus on the universal component of the entanglement entropy---specifically the corner correction encoding critical properties---which we denote as $\Delta S_2^A$. Motivated by the equilibrium scaling form in \Eq{scalingeq}, we propose a general ansatz for the corner contribution $\Delta S_2^A$ during imaginary-time evolution:
\begin{equation}
	\Delta S_2^A(L,\tau) = s_c(\theta) \ln L + \mathcal{F}(\tau/L^z) + \text{const.} + \mathcal{O}(1/L),
	\label{scalingeq1}
\end{equation}
where $z$ is the dynamic exponent of the quantum critical point (QCP) and $\mathcal{F}$ is a universal scaling function. The system relaxes to the equilibrium ground state in the limit $\tau \gg L^z$, for which the scaling behavior must recover~\Eq{scalingeq}. 

Conversely, in the short-time regime where $\tau \ll L^z$, we speculate that starting from a product state, the spread of entanglement is confined by a finite length scale $\xi \sim \tau^{1/z}$ and has not yet reached the system boundary. Consequently, $\Delta S_2^A(L,\tau)$ must scale with the correlation length $\xi$ rather than the subregion size $L$. This implies that the dependence on $L$ in~\Eq{scalingeq1} must vanish in the short-time regime. This physical requirement dictates the asymptotic behavior of the scaling function: $\mathcal{F}(x) \rightarrow \frac{s_c(\theta)}{z}\ln(x)$ as $x \rightarrow 0$. Therefore, in the early stages of evolution, we expect the universal corner contribution to obey the following scaling form:
\begin{equation}
	\Delta S_2^A(L,\tau) = \frac{s_c(\theta)}{z} \ln \tau + \text{const.} + \mathcal{O}(1/L).
	\label{scalingneq}
\end{equation}
Notably, in this regime, $\Delta S_2^A(L,\tau)$ is independent of the system size to leading order, as the correlation length remains much smaller than $L$. Hence, this short-time scaling form offers a powerful tool to extract the universal coefficient $s_c(\theta)$, which characterizes the critical properties of the QCP.

\begin{figure}[t]
	\centering
	\includegraphics[width=\linewidth]{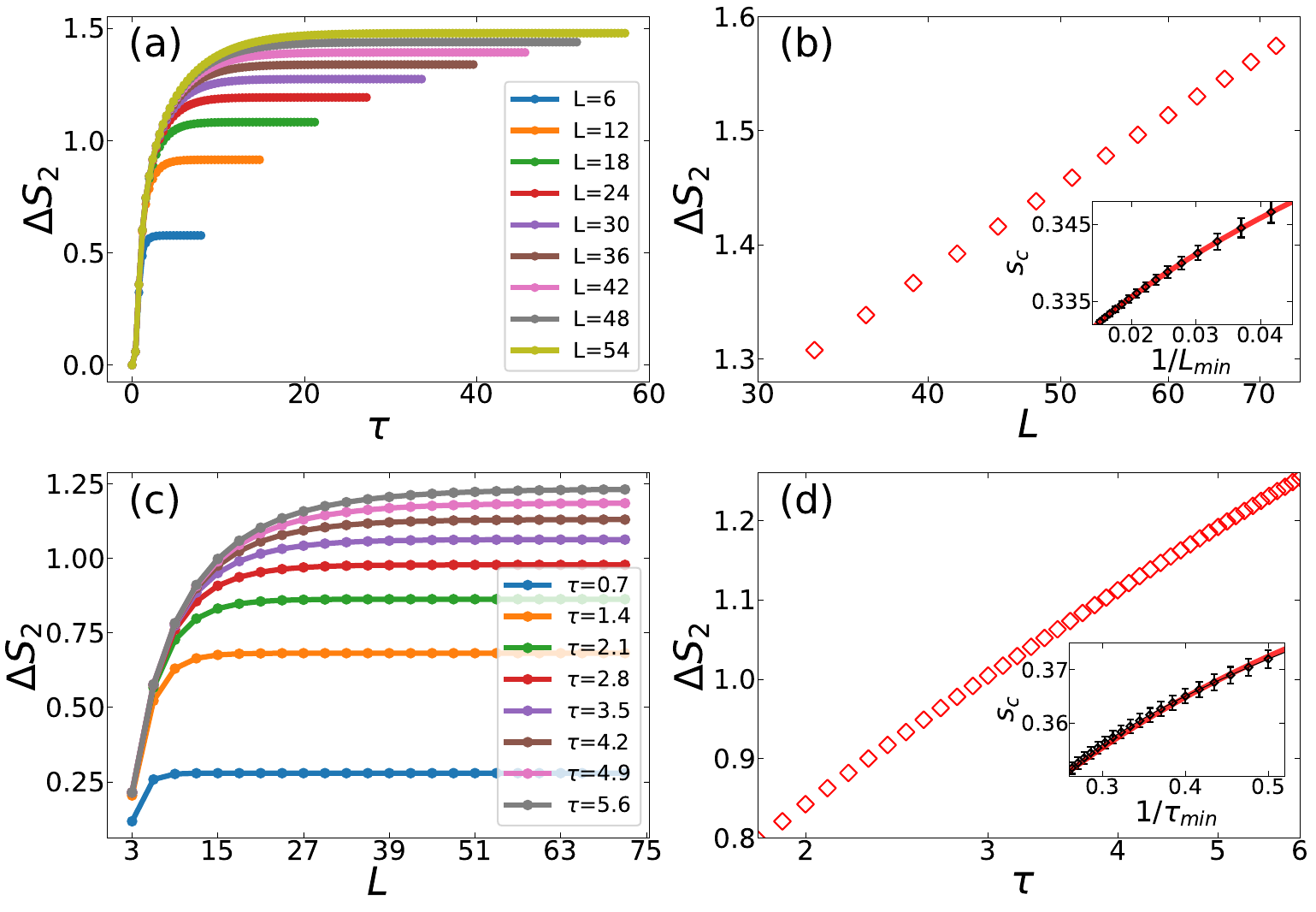}
	\caption{
		\textbf{Non-equilibrium relaxation and scaling of the corner entanglement entropy $\Delta S_2$ in the non-interacting ($U=0$) system.}
		\textbf{(a)} Imaginary-time evolution of $\Delta S_2$ from a product state toward the free Dirac fermion ground state. 
		\textbf{(b)} Scaling analysis of the equilibrium corner entanglement $\Delta S_2$ with system size $L$, demonstrating the logarithmic dependence of $\Delta S_2$ on $L$. We determine the equilibrium coefficient $s_c^{\mathrm{eq}}$ by fitting $\Delta S_2(L)$ to the form $s_c \ln L +\text{const.}$ over fitting windows with varying lower bounds $L_{\text{min}}$ and a fixed maximum $L_{\text{max}}=90$. Inset: Extrapolation of the fitting results of $s_c^{\mathrm{eq}}(L_{\text{min}})$ to $L_{\text{min}}^{-1}\rightarrow 0$ using a polynomial fit, yielding $s_c^{\mathrm{eq}} = 0.3116(13) $.
		\textbf{(c)} $\Delta S_2$ plotted against system size $L$ for different fixed values of imaginary time $\tau$. In the non-equilibrium relaxation regime ($\tau \ll L$), $\Delta S_2$ exhibits a ``size-independent plateau'' determined solely by $\tau$.
		\textbf{(d)} The universal size-independent scaling of $\Delta S_2$ plotted as a function of imaginary time $\tau$. Inset: We extract the coefficient by fitting $\Delta S_2(\tau)$ to the logarithmic form $s_c \ln \tau + \text{const.}$ in fitting windows with varying minima $\tau_{\text{min}}$ and a fixed $\tau_{\text{max}}=6.0$. Extrapolating to $\tau_{\text{min}}^{-1}\rightarrow 0$ yields $s_c^{\mathrm{neq}} = 0.3111(15)$. 
	}
	\label{U=0_neq}
\end{figure}

\begin{figure*}[t]
	\centering
	\includegraphics[width=\linewidth]{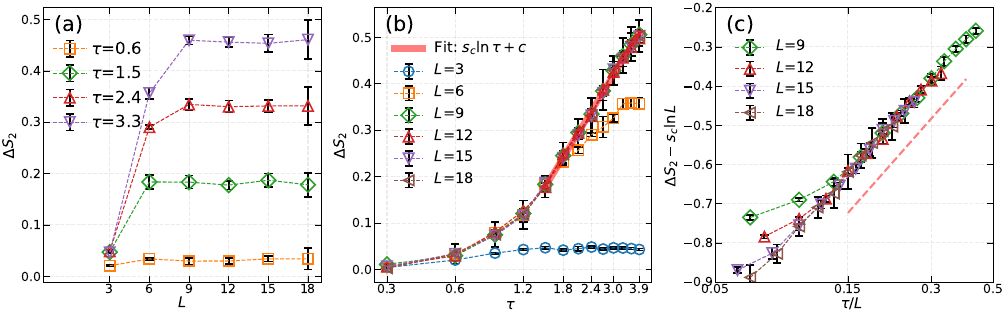}
	\caption{\textbf{Non-equilibrium dynamics of $\Delta S_2$ at the GNY QCP ($U = 3.8$).} 
		All simulations start from an AFM product state.
		\textbf{(a)} $\Delta S_2$ versus system size $L$ for different $\tau$. In the non-equilibrium regime ($\tau \ll L$), $\Delta S_2$ exhibits a size-independent plateau.
		\textbf{(b)} Logarithm plotting of $\Delta S_2$ versus $ \tau$. The data for different $L$ collapse and are fitted to $\Delta S_2 = s_c \ln \tau + \text{const.}$, yielding the non-equilibrium coefficient $s_c^{\text{neq}} = 0.345(7)$.
		\textbf{(c)} Data collapse validating the universal scaling form in Eq.~(\ref{scalingeq1}). Plotting $\Delta S_2 - s_c \ln L$ versus the scaling variable $\tau/L$ collapses all data onto a single master curve. The dashed line with the slope of $0.345$ is plotted for comparison with the rescaled curve. 
	}
	\label{U=3.8}
\end{figure*}

\noindent {\bf Model and Method}

We select a classic model of interacting fermions in two dimensions as the benchmark system for our new method: the Hubbard model on a honeycomb lattice at half filling. Its Hamiltonian is given by:
\begin{equation}
	\label{hamiltonian}
	\begin{split}
		H&=H_t+H_u\\
		&=-t\sum_{\avg{ij},\sigma}c_{i\sigma}^\dagger c_{j\sigma}+\frac{U}{2}\sum_i (n_{i\uparrow}+n_{i\downarrow}-1)^2,
	\end{split}
\end{equation}
in which $t$ is the hopping coefficient and $U$ represents the strength of the Hubbard interaction. The restriction to half filling ensures the absence of the notorious ``sign problem'' in QMC simulations~\cite{Sugar1990prb,Wu2005PRBsign,Li2015PRBsign,Li2016PRLsign,xiang2016prl,wang2015prl,Mondaini2022science,Yao2024arXivsignproblem}. This model is known to host a quantum phase transition from a Dirac semi-metal to an AFM insulator at a critical interaction strength $U_c \approx 3.8$~\cite{Assaad2013prx,Sorella2012SP,Sorella2016PRX}. In the AFM phase, the system undergoes spontaneous symmetry breaking from SU($2$) down to U($1$), resulting in $N_G=2$ Goldstone modes. Crucially, the effective field theory of model~(\ref{hamiltonian}) possesses Lorentz-invariant structure with a dynamic critical exponent $z=1$. This feature implies that the critical dynamics are governed by a relativistic CFT, where time and space scale identically ($\tau \sim L$).

We perform our calculations within the general framework of projector quantum Monte Carlo (PQMC) simulations~\cite{Sorella1989EPL,AssaadReview}. To achieve high numerical stability and precision, we employ an incremental algorithm for entanglement entropy~\cite{Emidio2024PRL,Liao2025npjQI,DaliaoYuan2023PRB}. 
We isolate the universal logarithmic term using the Subtracted Corner Entanglement Entropy (SCEE) method~\cite{kups7666,Xu2024PRLDisorderoperator,DaLiaoYuan2024PRB}, 
which calculates the difference $\Delta S_2 = S_2^B - S_2^A$ between two regions with identical boundary lengths but distinct geometries. 
Specifically, region $B$ has a smooth boundary ($s_c^B=0$), while region $A$ contains four sharp corners---two with $\theta = \pi/3$ and two with $\theta = 2\pi/3$ (see Fig.~\ref{SCEE_png}(a)). 
This subtraction directly isolates the corner contribution, where the universal coefficient $s_c$ in Eq.~\eqref{scalingeq} is given by $s_c=N_\mathrm{f/b}\times 2[a_2^\mathrm{f/b}(\frac{\pi}{3})+a_2^\mathrm{f/b}(\frac{2\pi}{3})]$, where $N_{\mathrm f}$ and $N_{\mathrm b}$ denote the number of fermion and boson species, respectively.
Here, $a_2^\mathrm{f/b}(\theta)$ represents the universal angle-dependent coefficients determined solely by the underlying CFT~\cite{Helmes2016PRB}. 
Further technical details are provided in the Supplementary Material (SM)~\cite{SM}.

\noindent {\bf Free Dirac-fermion systems}

Before analyzing the interacting critical point, we benchmark the non-equilibrium scaling using the exactly solvable non-interacting ($U=0$) Hubbard model. This system hosts four Dirac cones—two spin and two valley flavors—implying a total theoretical corner-entanglement coefficient $s_c=0.3116$, obtained by summing the QFT contributions $a_2^\mathrm{f}(\pi/3)= 0.0330$ and $a_2^\mathrm{f}(2\pi/3) =0.0059$~\cite{Helmes2016PRB}.

We initialize the imaginary-time evolution from an AFM product state, for which $\Delta S_2=0$. The subsequent relaxation of $\Delta S_2$ displays two distinct regimes. For fixed system size $L$, the quantity grows rapidly at short imaginary times and eventually saturates for $\tau \gg L$, yielding the ground-state value as shown in Fig. \ref{U=0_neq}\subref{(a)}. To extract the equilibrium coefficient, we fit the saturated values $\Delta S_2(L)$ as shown in the inset of Fig. \ref{U=0_neq}\subref{(b)} to the logarithmic form $\Delta S_2 = s_c \ln L + \text{const.}$. As shown in Fig. \ref{U=0_neq}\subref{(b)}, we use fitting windows $[L_{\min}, L_{\max}]$ with a fixed upper bound $L_{\max}=90$, and for each choice of $L_{\min}$, all system sizes $L \ge L_{\min}$ are included in the fit. The resulting size-dependent coefficients converge smoothly as $L_{\min}$ increases, producing the extrapolated equilibrium value $s_c^{\mathrm{eq}} = 0.3116(13) $ .

To probe the non-equilibrium scaling, we fix $\tau$ and study the dependence of $\Delta S_2$ on the system size in Fig. \ref{U=0_neq}\subref{(c)}. In the regime $\tau \ll L$, the results exhibit a clear size-independent plateau, consistent with the expectation that $\Delta S_2$ depends only on $\tau$ before the system relaxes to its equilibrium state Eq. (\ref{scalingneq}).  By plotting these plateau values in the inset of Fig. \ref{U=0_neq}\subref{(d)}, we can see a robust linear dependence for these plateau values versus $\ln \tau$. The finite-time coefficient is obtained by fitting $\Delta S_2(\tau) = s_c \ln \tau + \text{const.}$ over fitting windows $[\tau_{\min},\tau_{\max}]$ with a fixed upper bound $\tau_{\max}=6.0$, and extrapolating the extracted coefficients to $\tau_{\min}^{-1} \rightarrow 0$ as shown in Fig. \ref{U=0_neq}\subref{(d)}. This procedure yields $s_c^{\mathrm{neq}} = 0.3111(15)$, in excellent agreement with both the equilibrium result and the field-theoretic prediction.

Our non-equilibrium result, $s_c^{\mathrm{neq}}$, aligns closely with the CFT prediction $s_c = 0.3116$ for free Dirac fermions~\cite{Helmes2016PRB}. While both approaches suffer from finite-size effects, the non-equilibrium scaling converges significantly faster: stable universal coefficients are obtained with system sizes as small as $L=54$ for imaginary times $\tau \le 6.0$. In contrast, the equilibrium analysis requires substantially larger lattices (up to $L=90$) and corresponding long imaginary time ($\tau \sim L $) to achieve comparable precision. This demonstrates a practical advantage of the non-equilibrium method for extracting universal entanglement coefficients in 2D fermion systems.

\noindent {\bf Gross-Neveu-Yukawa quantum critical point}

We now extend this non-equilibrium protocol to the interacting GNY QCP, where the universal coefficient $s_c$ lacks a known analytical solution. While recent QMC studies have investigated $s_c$ at the GNY QCP~\cite{Emidio2024PRL,Liao2025npjQI}, those works focused on subsystem corners with angles of $\pi/3$ and $\pi/4$, respectively. In contrast, our setup utilizes a subsystem geometry on the honeycomb lattice that incorporates both $\pi/3$ and $2\pi/3$ corners, thereby probing a different set of universal corner contributions. Characterizing universal properties in this interacting regime via conventional equilibrium projector QMC is notoriously difficult due to critical slowing down. This typically demands prohibitive computational costs (e.g., $\tau \sim 2L$ for $L\ge18$) to reliably converge to the ground state in the thermodynamic limit ~\cite{Emidio2024PRL}.

Our method bypasses this limitation by exploiting the short-time relaxation dynamics from an AFM initial state, providing direct access to universal scaling with dramatically reduced resources (e.g., $L=9$ and $\tau<4$). The results at $U_c$ are summarized in Fig. \ref{U=3.8}. We observe the same sequence of scaling phenomena established in the non-interacting limit: $\Delta S_2$ shows a size-independent plateau governed by $\tau$ in the $L \ll \tau$ regime as shown in Fig. \ref{U=3.8}\subref{(a)}, and all data collapse when plotted against $\ln \tau$ in Fig. \ref{U=3.8}\subref{(b)}.

From the fit in Fig.~\ref{U=3.8}\subref{(b)}, we extract the non-equilibrium scaling coefficient: $s_c^{\text{neq}} = 0.345(7)$.  Crucially, in the non-equilibrium regime ($1 \lesssim \tau \ll L$), the size dependence of $\Delta S_2$ is negligible for $L \geq 9$. This enables the unambiguous extraction of the universal coefficient $s_c$ using only modest system sizes and short imaginary times. Our high-precision result for $s_c$ is significantly larger than the free-fermion prediction ($s_c = 0.3116$), a trend that aligns with recent findings in Ref.~\cite{Emidio2024PRL}. This result highlights the stronger infrared entanglement signatures produced by the coupling between fermions and critical bosons, distinguishing the interacting GNY criticality from its free-fermion counterpart.

Finally, we validate the full scaling hypothesis of Eq.~(\ref{scalingeq1}). As shown in Fig.~\ref{U=3.8}\subref{(c)}, plotting $\Delta S_2 - s_c\ln L$ versus $\tau/L$ results in the collapse of all data onto a single universal curve throughout the entire imaginary-time evolution regime $\tau \gtrsim 1$. In addition, in the semi-log scale, the slope of the rescaled curve, which just corresponding to the scaling function $\mathcal{F}(\tau/L^z)$, is close to $0.345$, Fig.~\ref{U=3.8}\subref{(c)}, confirming that $\mathcal{F}(\tau/L)\rightarrow s_c\ln(\tau/L)$ as $\tau<L$. These results validate the scaling ansatz and illustrate that the non-equilibrium relaxation timescale follows $\tau_{\text{neq}} \sim L$, as expected for a QCP with dynamical exponent $z=1$.


\begin{figure}[t]
	\centering
	\includegraphics[width=\linewidth]{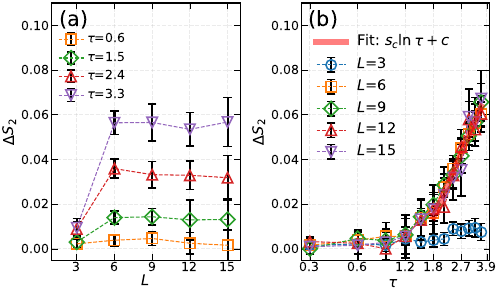}
	\caption{\textbf{Non-equilibrium dynamics of $\Delta S_2$ at the AFM phase ($U = 7$).} (a) $\Delta S_2$ versus $L$ for different $\tau$. In the non-equilibrium regime ($\tau \ll L$), $\Delta S_2$ exhibits a size-independent plateau. (b) Logarithm plotting of $\Delta S_2$ versus $\tau$.  The data for different $L$ collapse and are fitted to $\Delta S_2 = s_c \ln \tau + \text{const.}$, yielding the non-equilibrium coefficient $s_c^{\text{neq}} = 0.070(12) $.
	}
	\label{U=7}
\end{figure}

\noindent {\bf Antiferromagnetic phase}

In phases with spontaneously breaking of continuous symmetry, an additional logarithmic term related to the number of Goldstone modes ($N_G$) appears~\cite{SSB_scaling}: 
\begin{equation}
    S^A_2(L) = aL - (s_G + s_c(\theta)) \ln L + \text{const.} + \mathcal{O}(1/L),
\end{equation}
where $s_G = -N_G/2$. The phase under study is characterized by the spontaneous breaking of $SU(2)$ symmetry, resulting in $N_G = 2$ Goldstone modes (equivalent to two free real bosons). Notably, the $s_G$ term arises from the bulk properties of the symmetry-broken phase regardless of the boundary shape, while $s_c$ represents the specific corner contribution. By employing the SCEE subtraction scheme established above, we isolate $s_c$ by canceling out the $s_G$ term. For the antiferromagnetic (AFM) phase investigated here, each gapless boson is predicted to contribute approximately $0.0344$ to the logarithmic coefficient for our specific geometry~\cite{Helmes2016PRB}. Consequently, this model serves as an ideal benchmark for extracting universal corner contributions via our non-equilibrium approach.

To investigate the dynamics deep within the AFM Mott insulator phase and avoid critical fluctuations associated with the GNY QCP, we perform simulations at large interaction strength ($U=7$) starting from an AFM product state. The non-equilibrium dynamics exhibit scaling behavior qualitatively similar to the free fermion and QCP cases. As shown in Fig.~\ref{U=7}\subref{(a)}, in the non-equilibrium regime $\tau \ll L$, the system-size dependence of $\Delta S_2$ vanishes. In this regime, the corner contribution follows the scaling relation $\Delta S_2 \sim s_c \ln \tau$. Fitting this form yields a universal coefficient $s_c=0.070(12) $, as shown in Fig.~\ref{U=7}\subref{(b)}. Results for $U=6$ yield a consistent value of $s_c=0.069(10)$ (see SM). These findings are in excellent agreement with the theoretical prediction for two Goldstone modes ($2 \times 0.0344 = 0.0688$)~\cite{Helmes2016PRB}.

\noindent {\bf Discussion and conclusion}

We reveal a universal imaginary-time non-equilibrium scaling law for entanglement entropy in (2+1)D quantum critical systems. Through scaling analysis and numerical validation in free Dirac fermion systems, we establish that the corner entanglement contribution follows the general logarithmic relation $\Delta S_2 = s_c \ln \tau + \text{const.}$ within the non-equilibrium regime $\tau_0 < \tau \ll L$. Here, the coefficient $s_c$ encodes the universal properties of the underlying CFT. Remarkably, a defining feature of this regime is that the entanglement growth is effectively independent of system size, providing a direct window into the extraction of critical data characteristic of the thermodynamic limit.

These non-equilibrium  provides a powerful theoretical framework for extracting the universal entanglement properties that encode critical behavior at the QCP. 
Our protocol offers two decisive advantages over traditional equilibrium calculations.
First, it drastically reduces computational cost. 
In QMC simulations, complexity scales with imaginary time; notably, in the presence of the sign problem, this scaling becomes exponential~\cite{Troyer2005sign}. 
Therefore, our short-time approach is highly efficient. 
Second, because the leading term lacks size-dependent corrections, our approach allows for the extraction of universal coefficients using smaller system sizes compared to traditional equilibrium methods. 
Applying this technique, we extract the universal corner contribution coefficient of the entanglement entropy at the GNY QCP in the honeycomb Hubbard model---a value for which no analytical solution currently exists. Our high-precision  results reveal that the corner contribution at the GNY QCP is significantly enhanced relative to free fermions, highlighting the entanglement signatures of interacting fermions coupled to gapless bosons. This work opens a new avenue for efficiently calculating universal entanglement properties in exotic strongly correlated quantum systems, such as characterizing deconfined quantum critical points~\cite{Senthil_2004,Senthil2004_PRB,Sandvik2007,Zhao2022_PRL,Sandvik2024,Lin2024PRL} and Dirac spin liquids featuring emergent gauge fields~\cite{Balents2016ReviewQSL,Zhou2017RMP,Senthil2020ReviewQSL}. Furthermore, by exploiting the fact that the sign problem remains relatively weak during short-time evolution, our approach offers a viable route to studying entanglement at QCPs in interacting models that are typically plagued by severe sign problems.

Besides the significance in theoretical and numerical aspects, the non-equilibrium scaling form would provide a theoretical framework for analyzing entanglement on quantum simulators amd suggest immediate experimental applications. This progress is particularly timely; given the recent breakthroughs in realizing imaginary-time dynamics on superconducting quantum circuits, our proposed protocol can be directly implemented and verified on existing hardware~~\cite{Chan2020NP,Nishi2021npjQI,Pollmann2021PRXQuantum,Zhang2024PRBimaginarytime}.

\noindent {\bf Data availability}

The data that support the findings of this study are available from the corresponding authors (Z.X.L. and S.Y.) upon request.

\noindent {\bf Code availability}

All numerical codes in this paper are available upon request to the corresponding authors (Z.X.L. and S.Y.).

{\bf Acknowledgments}

C.Y.S and Z.X.L are supported by the National Natural Science Foundation of China under Grant Nos. 12347107 and 12474146, and Beijing Natural Science Foundation under Grant No. JR25007. S.Y. is supported by the National Natural Science Foundation of China (Grant No. 12222515), the Research Center for Magnetoelectric Physics of Guangdong Province (Grant No. 2024B0303390001), and the Guangdong Provincial Key Laboratory of Magnetoelectric Physics and Devices (Grant No. 2022B1212010008).

{\bf Author contributions}

Z.X.L. and S.Y. conceived the project and planned the study. C.Y.S. conducted the numerical simulations. All authors contributed to data analysis and the preparation of the manuscript.

\onecolumngrid
\newpage
\widetext
\thispagestyle{empty}

\setcounter{equation}{0}
\setcounter{figure}{0}
\setcounter{table}{0}
\renewcommand{\theequation}{S\arabic{equation}}
\renewcommand{\thefigure}{S\arabic{figure}}
\renewcommand{\thetable}{S\arabic{table}}

\pdfbookmark[0]{Supplementary Materials}{SM}
\begin{center}
    \vspace{3em}
    {\Large\textbf{Supplementary Materials for}}\\
    \vspace{1em}
    {\large\textbf{Universal Entanglement Growth along Imaginary Time in Quantum Critical Systems}}\\
    \vspace{0.5em}
\end{center}

\section{I. Projected Quantum Monte Carlo and Rényi Entropy}
\label{DQMC}

The Projected Quantum Monte Carlo (PQMC) method extracts the ground state $\ket{\Psi_g}$ from a trial wave function $\ket{\Psi_T}$ via imaginary-time evolution. Given a Hamiltonian $H = H_0 + H_U$, the projection is defined as:
\begin{equation}
    \ket{\Psi_g} = \lim_{\tau \to \infty} e^{-\tau H} \ket{\Psi_T}.
\end{equation}
To evaluate the imaginary-time propagator, we discretize the total projection time $2\tau$ into $N_\tau$ slices of width $\Delta_\tau = 2\tau / N_\tau$. Since the non-interacting ($H_0$) and interacting ($H_U$) terms do not commute, we employ a first-order Trotter-Suzuki decomposition to approximate the partition function:
\begin{equation}
    \mathcal{Z} = \bra{\Psi_T} e^{-2\tau H} \ket{\Psi_T} = \bra{\Psi_T} \left( e^{-\Delta_\tau H_0} e^{-\Delta_\tau H_U} \right)^{N_\tau} \ket{\Psi_T} + \mathcal{O}(\Delta_\tau^2).
\end{equation}
Because $H_U$ contains quartic fermionic operators, we decouple the interaction using a discrete $SU(2)$ Hubbard-Stratonovich (HS) transformation. By coupling the auxiliary fields $s_i$ to the local charge density, the interacting propagator is expressed as:
\begin{equation}
    e^{-\Delta_\tau \frac{U}{2}(n_{i,\uparrow}+n_{i,\downarrow}-1)^{2}} = \frac{1}{4} \sum_{s_{i} \in \{\pm 1, \pm 2\}} \gamma(s_{i}) e^{i \alpha \eta(s_{i}) (n_{i,\uparrow}+n_{i,\downarrow}-1)} + \mathcal{O}(\Delta_{\tau}^{4}),
\end{equation}
where $\alpha = \sqrt{\Delta_\tau U/2}$ and the constants are given by $\gamma(\pm 1) = 1 + \sqrt{6}/3$, $\gamma(\pm 2) = 1 - \sqrt{6}/3$, $\eta(\pm 1) = \pm\sqrt{2(3-\sqrt{6})}$, and $\eta(\pm 2) = \pm\sqrt{2(3+\sqrt{6})}$.

Employing the Trotter-Suzuki decomposition and the discrete four-point Hubbard-Stratonovich (HS) transformation, the projection operator is expressed as a summation over auxiliary field configurations $\mathbf{s}$. This formulation enables the Monte Carlo sampling of the partition function $\mathcal{Z} = \langle \Psi_T | e^{-2\tau H} | \Psi_T \rangle$~\cite{AssaadReview}. To control discretization errors, we set the imaginary time step to $\Delta \tau = 0.1$ for interaction strengths $U = 3.8$ and $\Delta \tau = 0.03$ for $U = 6, 7$. The partition function is given by:
\begin{equation}
    \mathcal{Z} = \sum_{\mathbf{s}} \mathcal{W}_{\mathbf{s}}, \quad 
    \mathcal{W}_{\mathbf{s}} = \det \{ \mathbf{P}^{\dagger} \mathbf{B}_{\mathbf{s}}(2\tau, 0) \mathbf{P} \},
\end{equation}
where $\mathbf{P}$ is the matrix representation of the trial wavefunction $| \Psi_T \rangle$, and $\mathbf{B}_{\mathbf{s}}$ encodes the propagation of the kinetic term $H_0$ coupled with the auxiliary-field potential $V(\mathbf{s}_l)$.

A cornerstone of the PQMC framework is the equal-time Green's function at a specific imaginary time layer $\Theta$:
\begin{equation}
    \mathbf{G}_{\mathbf{s}}(\Theta) = \mathbb{I} - \mathbf{B}_{\mathbf{s}}(\tau, 0)\mathbf{P} \left[ \mathbf{P}^{\dagger}\mathbf{B}_{\mathbf{s}}(2\tau, 0)\mathbf{P} \right]^{-1} \mathbf{P}^{\dagger}\mathbf{B}_{\mathbf{s}}(2\tau, \Theta).
\end{equation}
The corresponding time-displaced Green's function $\mathbf{G}_{\mathbf{s}}(\tau_1, \tau_2)$ is defined as:
\begin{equation}
    \mathbf{G}_{\mathbf{s}}(\tau_1, \tau_2) =
    \begin{cases}
        -[\mathbb{I} - \mathbf{G}_{\mathbf{s}}(\tau_1)] \mathbf{B}_{\mathbf{s}}^{-1}(\tau_2, \tau_1), & \tau_1 < \tau_2, \\
        \mathbf{B}_{\mathbf{s}}(\tau_1, \tau_2) \mathbf{G}_{\mathbf{s}}(\tau_2), & \tau_1 > \tau_2.
    \end{cases}
\end{equation}

The second Rényi entanglement entropy, $S_2^A$, is evaluated using the identity 
$e^{-S_2^A} = \text{Tr}[\rho_A^2] = Z_2/Z_1^2 = \langle \det \mathbf{g}^A \rangle$. 
The Grover matrix $\mathbf{g}^A$, restricted to subsystem $A$, is constructed from 
the Green's functions of two independent replicas, $\mathbf{s}_1$ and $\mathbf{s}_2$, 
at the imaginary-time midpoint $\tau$~\cite{Grover2013PRL}:
\begin{equation}
    \mathbf{g}^A_{\mathbf{s}_1, \mathbf{s}_2} = \mathbf{G}^A_{\mathbf{s}_1}(\tau)\mathbf{G}^A_{\mathbf{s}_2}(\tau) + [\mathbb{I} - \mathbf{G}^A_{\mathbf{s}_1}(\tau)][\mathbb{I} - \mathbf{G}^A_{\mathbf{s}_2}(\tau)].
\end{equation}
In the Monte Carlo sampling, the acceptance ratio for a single auxiliary-field update is given by
\begin{equation}
    r = \prod_{\sigma} \left[ \mathbb{I} + \Delta_{x}^{\sigma} (\mathbb{I} - \mathbf{G}_{\mathbf{s}}^{\sigma}(\Theta)) \right],
\end{equation}
where $\Delta_{x}^{\sigma}$ denotes the rank-1 update matrix. To maintain computational efficiency, 
the imaginary-time-displaced Green's functions are updated following each accepted 
configuration change according to~\cite{DaliaoYuan2023PRB}:
\begin{equation}
    \left\{
    \begin{aligned}
        \mathbf{G}_{\mathbf{s}_{1}^\prime}(\Theta) &= \mathbf{G}_{\mathbf{s}_{1}}(\Theta) - \mathbf{G}_{\mathbf{s}_{1}}(\Theta)\mathbf{r}^{-1}\Delta[\mathbb{I}-\mathbf{G}_{\mathbf{s}_{1}}(\Theta)] ,\\
        \mathbf{G}_{\mathbf{s}_1^\prime}(\tau) &= \mathbf{G}_{\mathbf{s}_1}(\tau) + \mathbf{G}_{\mathbf{s}_1}(\tau,\Theta)\mathbf{r}^{-1}\Delta \mathbf{G}_{\mathbf{s}_1}(\Theta,\tau), \\
        \mathbf{G}_{\mathbf{s}_1^\prime}(\Theta,\tau) &= \mathbf{G}_{\mathbf{s}_1}(\Theta,\tau) + \mathbf{G}_{\mathbf{s}_1}(\Theta)\mathbf{r}^{-1}\Delta \mathbf{G}_{\mathbf{s}_1}(\Theta,\tau), \\
        \mathbf{G}_{\mathbf{s}_{1}^\prime}(\tau,\Theta) &= \mathbf{G}_{\mathbf{s}_{1}}(\tau,\Theta) - \mathbf{G}_{\mathbf{s}_{1}}(\tau,\Theta)\mathbf{r}^{-1}\Delta[\mathbb{I}-\mathbf{G}_{\mathbf{s}_{1}}(\Theta)].
    \end{aligned}
    \right.
\end{equation}
Similarly, the Grover matrix $\mathbf{g}^A_{\mathbf{s}_1, \mathbf{s}_2}$ can be updated efficiently. 
An update in the configuration of the first replica, $\mathbf{s}_1 \to \mathbf{s}_1^\prime$, 
results in the relation $\mathbf{g}^A_{\mathbf{s}_1^\prime, \mathbf{s}_2} (\mathbf{g}^A_{\mathbf{s}_1, \mathbf{s}_2})^{-1} = \mathbb{I} + \hat{a} \hat{b}$, with:
\begin{equation}
    \hat{a} = \mathbf{G}_{\mathbf{s}_1}(\tau, \Theta) \mathbf{r}^{-1}\Delta, \quad 
    \hat{b} = \mathbf{G}_{\mathbf{s}_1}(\Theta, \tau) [2\mathbf{G}_{\mathbf{s}_2}(\tau) - \mathbb{I}] (\mathbf{g}^{A}_{\mathbf{s}_1, \mathbf{s}_2})^{-1}.
\end{equation}
An analogous expression holds for updates to the second replica, $\mathbf{s}_2 \to \mathbf{s}_2^\prime$, where:
\begin{equation}
    \hat{a} = [2\mathbf{G}_{\mathbf{s}_1}(\tau) - \mathbb{I}] \mathbf{G}_{\mathbf{s}_2}(\tau, \Theta), \quad 
    \hat{b} = \mathbf{r}^{-1}\Delta \mathbf{G}_{\mathbf{s}_2}(\Theta, \tau) (\mathbf{g}_{\mathbf{s}_1, \mathbf{s}_2}^{A})^{-1}.
\end{equation}
By exploiting the rank‑1 structure of $\Delta$, only a single row or column slice of the matrices $\hat{a}$ and $\hat{b}$ is required to enable efficient and rapid updates to the Grover matrix $\mathbf{g}^{A}$.

\section{II. Incremental Algorithm for Entanglement Entropy}
\label{Incremental}

The Rényi entropy $S_2$ exhibits extensive scaling with subsystem size. Although Grover's method within the DQMC framework provides an unbiased estimator for $e^{-S_2}$~\cite{Grover2013PRL}, the exponential decay of this quantity with system size $L$ results in a severe signal-to-noise ratio problem. To mitigate the exponentially growing relative statistical errors, we employ the incremental algorithm. We decompose the expectation value $O \equiv \det \mathbf{g}^A$ into a product of incremental ratios along a monotonically increasing path $\boldsymbol{\lambda} = \{\lambda_0, \lambda_1, \dots, \lambda_N\}$ with $\lambda_0=0$ and $\lambda_N=1$~\cite{DaliaoYuan2023PRB,  Emidio2024PRL}:
\begin{equation}
    \langle O \rangle = \prod_{n=0}^{N-1} \frac{\sum_i P_i O_i^{\lambda_{n}} O_i^{\Delta_\lambda}}{\sum_i P_i O_i^{\lambda_{n}}}.
\end{equation}

\begin{figure}[!t]
    \centering
    \includegraphics[width=\linewidth]{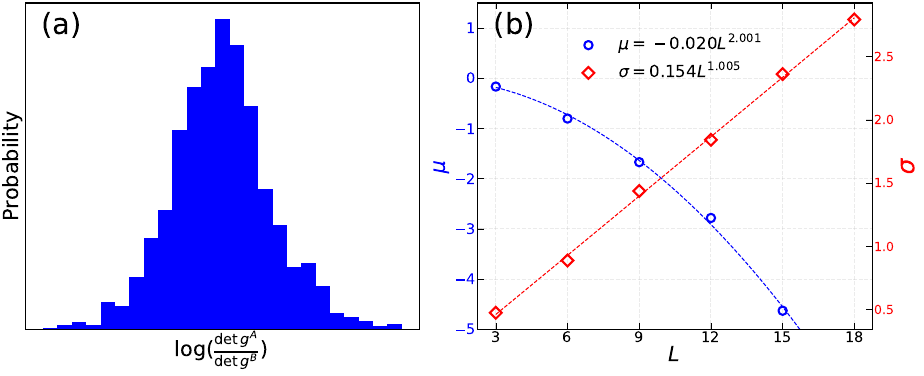}
    \caption{(a) Sampling distribution of $\log(\det \mathbf{g}^A / \det \mathbf{g}^B)$, which follows a normal distribution $\mathcal{N}(\mu, \sigma^2)$. (b) Scaling of the statistical mean $\mu$ (left) and standard deviation $\sigma$ (right) with system size $L$. The solid lines represent fits to the scaling relations $\mu \approx -0.02L^{1.995}$ and $\sigma \approx 0.165L^{0.981}$.}
    \label{Nicr}
\end{figure}

For systematic analysis, we utilize a linear parameterization $\lambda_n = n/N_\lambda$, where $\Delta_{\lambda_n} = 1/N_\lambda$ is the incremental step size. As the lattice size $L$ increases, the number of intermediate $\lambda$-points must be increased to constrain the statistical error. As illustrated in Fig.~\ref{Nicr}(a), the logarithm of the determinant, $\ln(\det \mathbf{g}^A)$, follows a normal distribution $\mathcal{N}(\mu, \sigma^2)$. By fitting the data (Fig.~\ref{Nicr}(b)), we extract the scaling behavior of the mean $\mu(L) \propto L^\alpha$ and standard deviation $\sigma(L) \propto L^\beta$. 

Crucially, the coefficient of variation for the raw determinant, $CV[\det \mathbf{g}^A] = \sqrt{e^{\sigma^2}-1}$, grows exponentially with $L$. In contrast, the coefficient of variation for the incremental step scales as $CV[(\det \mathbf{g}^A)^{1/N_\lambda}] = \sqrt{e^{\sigma^2/N_\lambda^2}-1}$. Consequently, to ensure numerical precision, we select $N_\lambda(L) \approx 0.2L^{1.7} > \sigma(L)$. Specifically, we use $N_\lambda = \{2, 5, 9, 15, 21, 28\}$ for system sizes $L = \{3, 6, 9, 12, 15, 18\}$, respectively.

At each step $n$, the effective weight and observable are defined as:

\begin{equation}
    \left\{
    \begin{aligned}
        &\mathbf{P}_{i}^{(n)} \equiv P_i O_i^{\lambda_{n}}, \\
        &O_{i}^{(n)} \equiv O_i^{\Delta_{\lambda_n}}.
    \end{aligned}
    \right.
\end{equation}
The modified acceptance ratio for a single auxiliary field flip is $R_{\lambda_n} = r \cdot (\det \Gamma_A)^{\lambda_n}$, where $r$ is the standard PQMC update ratio and $\Gamma_A \equiv \mathbb{I} + \hat{b}  \hat{a}$ accounts for the subsystem contribution.

\section{III. Subtracted Corner Entanglement Entropy}
\label{SCEE}
The universal corner contribution to the second Rényi entropy is isolated through the subtracted corner entanglement entropy $\Delta S_2 = S_2^A - S_2^B$. Here, $S_2^A$ and $S_2^B$ denote the entanglement entropies for subsystems with geometries $A$ and $B$, respectively, which are carefully constructed to share identical boundary lengths but differ only in the presence of corners. This subtraction scheme effectively cancels the non-universal area-law contributions, leaving only the universal corner term. To circumvent numerical instabilities and the prohibitive computational overhead inherent in independent estimations of $S_2^A$ and $S_2^B$, we directly sample the exponential difference $e^{-\Delta S_2}$ through a ratio of determinants~\cite{kups7666, Xu2024PRLDisorderoperator, DaLiaoYuan2024PRB}:
\begin{equation}
    e^{-(S_A-S_B)} = \frac{\sum_{\mathbf{s}_1,\mathbf{s}_2} P_{\mathbf{s}_1,\mathbf{s}_2} \det \mathbf{g}^B_{\mathbf{s}_1,\mathbf{s}_2} \cdot \frac{\det \mathbf{g}^A_{\mathbf{s}_1,\mathbf{s}_2}}{\det \mathbf{g}^B_{\mathbf{s}_1,\mathbf{s}_2}}}{\sum_{\mathbf{s}_1,\mathbf{s}_2} P_{\mathbf{s}_1,\mathbf{s}_2} \det \mathbf{g}^B_{\mathbf{s}_1,\mathbf{s}_2}},
\end{equation}
where $\mathbf{g}^{A/B}$ are the Grover matrices restricted to subsystems $A$ and $B$, constructed from the mid-time Green's functions of two independent auxiliary field configurations $\mathbf{s}_1, \mathbf{s}_2$. In the combined Incremental and Subtracted (ICR-SCEE) method, we use the linear ramp of incremental path $\lambda_n \in [0,1]$ to ensure the accuracy as:
\begin{equation}
    \left\{
    \begin{aligned}
        \mathcal{P}_{i}^{(n)} &\equiv P_{\mathbf{s}_1,\mathbf{s}_2} (\det \mathbf{g}^B_{\mathbf{s}_1,\mathbf{s}_2})^{1-\lambda_n} (\det \mathbf{g}^A_{\mathbf{s}_1,\mathbf{s}_2})^{\lambda_n}, \\
        \mathcal{O}_{i}^{(n)} &\equiv \left( \frac{\det \mathbf{g}^A_{\mathbf{s}_1,\mathbf{s}_2}}{\det \mathbf{g}^B_{\mathbf{s}_1,\mathbf{s}_2}} \right)^{\Delta_{\lambda_n}}.
    \end{aligned}
    \right.
\end{equation}
This formulation ensures a smooth interpolation between the geometries of subsystems $A$ and $B$. The corresponding ICR-SCEE update ratio for an auxiliary field flip is:
\begin{equation}
    R_{\lambda_n} = r \cdot \det(\Gamma_A)^{\lambda_n} \cdot \det(\Gamma_B)^{1-\lambda_n},
\end{equation}
where $r$ is the standard PQMC update ratio, and $\Gamma_{A/B}$ represent the subsystem-dependent update factors for $\det \mathbf{g}^{A/B}$ derived from rank-1 matrix modifications.

\section{IV. Benchmark to the results of free Dirac fermions and gapless bosons}

\begin{figure}[t]
	\centering
	\includegraphics[width=\linewidth]{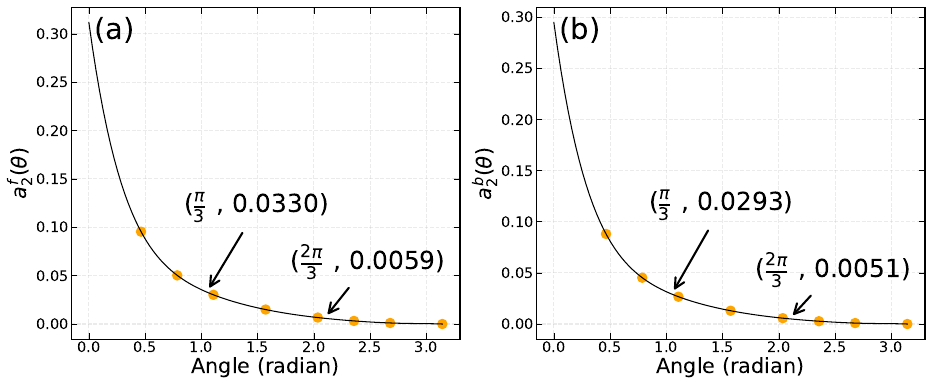}
	\caption{Corner contributions to the second R\'enyi entropy ($\alpha=2$) for (a) Dirac fermions and (b) free bosons. The numerical data points correspond to the high-precision results reported in Ref.~\cite{Helmes2016PRB}. To obtain the corner function values at $\theta = \pi/3$ and $2\pi/3$ required for our analysis, we perform a controlled polynomial interpolation of the tabulated data. This method preserves the smooth angular dependence observed in the corner function, enabling precise evaluation of $a_2^{\mathrm{f}/\mathrm{b}}(\pi/3)$ and $a_2^{\mathrm{f}/\mathrm{b}}(2\pi/3)$.}
	\label{BenchMark}
\end{figure}

To provide a rigorous validation of our numerical results, we benchmark the corner contribution to the second Rényi entropy, $s_c$, against established values for free relativistic theories. In the low-energy limit, the Dirac semi-metal and the antiferromagnetic (AFM) phases of our model are described by free Dirac fermions and gapless relativistic bosons, respectively. Given that the angles accessible in our calculations generally do not align with the discrete set of angles tabulated in Ref.~\cite{Helmes2016PRB}, we perform a controlled polynomial interpolation of the published numerical data, as illustrated in Fig.~\ref{BenchMark}. This interpolation method ensures continuity with the smooth-angle dependence observed in the corner function, enabling precise evaluation of $a_2^{\mathrm{f}/\mathrm{b}}(\pi/3)$ and $a_2^{\mathrm{f}/\mathrm{b}}(2\pi/3)$. Here, $a_2^{\mathrm{f}/\mathrm{b}}(\theta)$ denotes the universal coefficient of the logarithmic corner correction to the R\'{e}nyi entropy $S_2$ for a single flavor of Dirac fermions (f) or gapless relativistic complex bosons (b), respectively.

The honeycomb lattice hosts four flavors of degenerate gapless Dirac fermions, accounting for both spin and valley degrees of freedom. In the Dirac semi-metal phase, the total corner contribution $s_c$ is obtained by summing the contributions from all flavors across the relevant opening angles:
\begin{equation}
    s_c = 4 \times 2 \left[ a_2^{\mathrm{f}}(\pi/3) + a_2^{\mathrm{f}}(2\pi/3) \right] = 0.3116.
\end{equation}
Here, the factor of 4 accounts for the four fermion flavors, while the factor of 2 represents the two pairs of corners (with angles $\pi/3$ and $2\pi/3$) present in our entanglement geometry.

In the antiferromagnetic (AFM) phase, the low-energy physics is governed by two Goldstone modes corresponding to two real free bosons (equivalent to one complex free boson). The associated corner contributions are given by
\begin{equation}
\frac{1}{2} \times 2 \left[ a_2^{\mathrm{b}}(\pi/3) + a_2^{\mathrm{b}}(2\pi/3) \right] = 0.0344.
\end{equation}
The prefactor $\frac{1}{2}$ accounts for the equivalence between two real bosons and one complex boson, while the summation captures the angular dependence of the corner function for each Goldstone mode.

\section{V. Additional result of quantum monte-carlo simulation}

To corroborate the universality of our non-equilibrium protocol within the symmetry-broken phase, we extended our projector quantum Monte Carlo (PQMC) simulations to an interaction strength of $U=6$. This parameter regime lies deep within the antiferromagnetic (AFM) Mott insulating phase but possesses distinct microscopic energy scales compared to the $U=7$ case presented in the main text. In the AFM phase, the continuous $SU(2)$ spin rotational symmetry is spontaneously broken down to $U(1)$, resulting in $N_G=2$ gapless Goldstone modes (spin waves). These modes are expected to dominate the universal corner entanglement contribution, theoretically predicted to be $s_c = 0.0688$, arising from the sum of two free bosonic modes ($2 \times 0.0344$)~\cite{Helmes2016PRB}.

\begin{figure}[t]
	\centering
	\includegraphics[width=\linewidth]{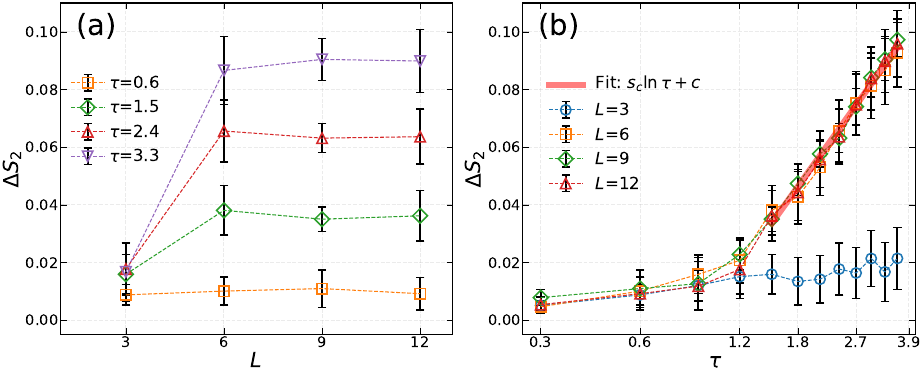}
	\caption{Non-equilibrium dynamics of $\Delta S_2$ at $U = 6$. (a) $\Delta S_2$ vs. system size $L$ for various $\tau$, showing size-independent plateau. (b) Logarithmic plot of $\Delta S_2$ vs. $\tau$, revealing linear scaling in non-equilibrium regime. The extracted corner contribution is $s_c = 0.069(10)$. }
	\label{U=6}
\end{figure}

As illustrated in Fig.~\ref{U=6}\subref{(a)}, the subtracted corner entanglement entropy $\Delta S_2$ exhibits the characteristic size-independent plateau in the short-time relaxation regime ($\tau \ll L$). This confirms that, the short-time dynamics are governed by local entanglement spreading rather than finite-size geometry. By performing a logarithmic fit to the scaling ansatz $\Delta S_2(\tau) \sim s_c \ln \tau$ over the temporal window corresponding to this plateau, as shown in Fig.~\ref{U=6}\subref{(b)}, we extract a non-equilibrium universal coefficient of $s_c = 0.069(10)$.

These results are significant physical consequence for two primary reasons. First, the value obtained is consistent with $s_c = 0.070(12)$ measured at a stronger coupling ($U=7$) within error bars. Second, it demonstrates excellent quantitative agreement with the rigorous field-theoretic prediction for free relativistic bosons. The stability of $s_c$ across a range of Hubbard repulsions $U$ serves as a robust validation of our methodology. 
It confirms that the imaginary-time dynamics effectively filter out non-universal microscopic details, providing a reliable and efficient probe for the universal entanglement features of phases characterized by the spontaneous breaking of continuous symmetries.

\end{document}